\documentclass[adp,a4paper,fleqn%
]{w-art}
\usepackage{times,cite,w-thm}
\theoremstyle{plain}

\theoremstyle{definition}

\usepackage[]{graphicx}
\usepackage{amsmath}
\usepackage{amssymb}
\begin{document}
%%    The information for the title page will be placed between
%%    \begin{document} and \maketitle. The order of most entries
%%    is determined by the class file and can not be changed by
%%    rearranging them. The maketitle command follows after the
%%    abstract.
%%
%%    Most of the following commands will be completed by the publisher.
%%
%%    The copyrightyear is defined in the .clo file as the first argument
%%    of the copyrightinfo command. If the copyrightyear differs from that
%%    value it might be adjusted by the following definition:
%%
%% \renewcommand{\copyrightyear}{2007}% uncomment to change the copyrightyear.
%%
\DOIsuffix{theDOIsuffix}
%%
%% issueinfo for the header line
\Volume{16}
\Month{01}
\Year{2007}
%%
%%    First and last pagenumber of the article. If the option
%%    'autolastpage' is set (default) the second argument may be left empty.
\pagespan{1}{}
%%
%%    Dates will be filled in by the publisher. The 'reviseddate' and
%%    'dateposted' (Published online) entry may be left empty.
\Receiveddate{XXXX}
\Reviseddate{XXXX}
\Accepteddate{XXXX}
\Dateposted{XXXX}

%% \pretitle{Editor's Choice}

%% We have a short and a long form for the title. The short form
%% (optional argument) goes into the running head.

\title[Multiple Dark Matter and dark interactions]{Multiple Dark Matter as a self-regulating mechanism for dark sector interactions}

%% Please do not enter footnotes or \inst{}-notes into the optional
%% argument of the author command. The optional argument will go into
%% the header.  If there is only one address the marker \inst{x} may be
%% omitted.

%% Information for the first author.
\author[M. Baldi]{Marco Baldi\inst{1,}%
  \footnote{\quad E-mail:~\textsf{marco.baldi@universe-cluster.de},
            %Phone: +00\,999\,999\,999,
            %Fax: +00\,999\,999\,999}
            }
\address[\inst{1}]{Excellence Cluster Universe,
Boltzmannstrasse 2, 85748 - Garching bei M\"unchen, Germany}
%%
%%    Information for the second author
%\author[S. Author]{Second Author\inst{1,2,}\footnote{Second author footnote.}}
%\address[\inst{2}]{Second address}
%%
%%    Information for the third author
%\author[T. Author]{Third Author\inst{2,}\footnote{Third author footnote.}
}
%%
%%    \dedicatory{This is a dedicatory.}
\begin{abstract}
A wide range of astrophysical and cosmological observations support
the evidence that the energy density of the Universe is presently largely dominated by 
particles and fields that do not belong to the standard model of particle physics. Such 
cosmic {\em dark sector} appears to be made of two distinct entities capable to account
for the growth of large-scale structures and for the observed acceleration of the expansion
rate of the Universe, respectively dubbed dark matter and dark energy. Nevertheless, the fundamental
nature of these two dark components has so far remained mysterious.
In the currently accepted scenario dark matter is associated to a single new massive and weakly interacting
particle beyond the standard model, while dark energy is assumed to be a simple cosmological constant.
However, present cosmological constraints and the absence of a direct detection and identification
of any dark matter particle candidate leave room to the possibility that the dark sector of the Universe
be actually more complex than it is normally assumed. In particular, more than one new fundamental
particle could be responsible for the observed dark matter density in the Universe, and possible new
interactions between dark energy and dark matter might characterize the dark sector. In the present work,
we investigate the possibility that two dark matter particles exist in nature, with identical physical properties except for the sign of their coupling constant to dark energy. Extending previous works on similar
scenarios, we study the evolution of the background cosmology as well as the growth of linear density
perturbations for a wide range of parameters of such model. Interestingly, our results show how the simple assumption that dark matter particles carry a {\em ``charge"} with respect to
their interaction with the dark energy field allows for new long-range scalar forces of gravitational strength in the dark sector without conflicting with present observations both at the background and linear levels. Our scenario does not introduce new parameters with respect to the case of a single dark matter species for which such strong dark interactions have been already ruled out. Therefore, the present investigation suggests that only a detailed study of nonlinear structure formation processes might possibly provide effective constraints on new scalar interactions of gravitational strength in the dark sector.
\end{abstract}
%% maketitle must follow the abstract.
\maketitle                   % Produces the title.

%% If there is not enough space inside the running head
%% for all authors including the title you may provide
%% the leftmark in one of the following three forms:

%% \renewcommand{\leftmark}
%% {First Author: A Short Title}

%% \renewcommand{\leftmark}
%% {First Author and Second Author: A Short Title}

%% \renewcommand{\leftmark}
%% {First Author et al.: A Short Title}

%% \tableofcontents  % Produces the table of contents.
\section{Introduction}

As most of the readers of these words will certainly know, the concept of dark matter has been introduced for the first time
in 1937 by the swiss astronomer Fritz Zwicky to indicate the missing mass required to explain the observed motion of galaxies
in the Coma cluster \cite{Zwicky_1937}. However, at the time of Zwicky such missing mass could have been assumed to be hidden
in galaxy clusters in the form of some invisible fraction of standard baryonic matter. The issue of the missing mass was therefore
more an open problem for observational astronomers rather than an indication of a failure of fundamental physics.
However, with the development of observational cosmology it has become progressively more evident that a significant amount of matter
with a completely different fundamental nature with respect to particles belonging to the standard model of particle physics must be
present in the Universe. This result was first suggested by the observed relative abundance of light elements in the Universe, 
that according to the predictions of Big Bang Nucleosynthesis puts tight constraints on the total cosmic baryonic density \cite{Reeves_etal_1973,Epstein_Lattimer_Schramm_1976,Schramm_Turner_1998}. Such observation, in combination with the determination of 
the total matter density in the Universe as inferred from complementary probes (e.g. \cite{SNLS,Borgani_etal_2001,Hoekstra_Yee_Gladders_2002, Holder_Haiman_Mohr_2001, Grego_etal_2001,Turner_2001} and references therein), provides compelling evidence of the existence
of a large amount of matter in the form of particles that do not belong to the standard model of particles physics.
The development of large N-body simulations has then allowed to investigate with ever increasing detail the properties of cosmic structures at different scales and provided a direct way to test the Cold Dark Matter paradigm with a wide range of observational techniques (see e.g. \cite{Davis_etal_1985,Navarro_Frenk_White_1995,Jenkins_etal_1998,Millennium,Aquarius,Angulo_etal_2012}).

At the present time, the existence of non-baryonic dark matter in the Universe is supported by a large number of independent and complementary
data, ranging from the anisotropies of the Cosmic Microwave Background (CMB, see e.g. \cite{wmap5,wmap7} and references therein)
to the formation and evolution of the cosmic Large Scale Structures (LSS, e.g. \cite{Percival_etal_2001,Reid_etal_2010}), from
the dynamical and thermodynamical properties of galaxies and galaxy clusters \cite{Persic_Salucci_1988,Reiprich_Boehringer_2002,Sand_etal_2004,Vikhlinin_etal_2006} to the lensing patterns of distant sources \cite{Kaiser_1992,Bartelmann_1996,Bacon_Refregier_Ellis_2000,Meneghetti_etal_2001,Hoekstra_Yee_Gladders_2004,Sand_etal_2002,Fedeli_etal_2008,Amendola_Kunz_Sapone_2008,Fu_etal_2008} or
to the study of colliding astrophysical systems such as the ``Bullet Cluster" \cite{Markevitch_etal_2001,Markevitch_etal_2003}.
However, all such probes
infer the existence of dark matter from its gravitational effects at galactic and extragalactic scales (up to its cosmological implications) and constrain its microscopical properties from the internal structure of large astrophysical objects,
while a direct detection and identification of possible fundamental
dark matter particle candidates has so far eluded any experimental effort. 
In the absence of a clear identification, it is therefore impossible to
constrain the fundamental nature of dark matter even though a number of well-motivated candidates from particle physics theories beyond the standard model have been proposed, such as
the {\em neutralino} in the context of supersymmetry \cite{Ellis_Olive_2010,Bertone_Hooper_Silk_2005} or the {\em axion} from theories aimed to solve the strong CP problem of QCD \cite{Preskill_Wise_Wilczek_1983}. In particular, since the present observational
evidence for dark matter is based mostly on its gravitational effects at galactic and extragalactic scales, it is not
possible to exclude a higher level of complexity of the dark matter sector at the microscopic level, as e.g. the possibility that more than one fundamental particle contribute to the
overall dark matter density. 

The situation has become even more entangled after the discovery that the Universe must be presently dominated by some other unknown field capable to
drive the observed accelerated expansion \cite{Riess_etal_1998,Perlmutter_etal_1999,Schmidt_etal_1998}, which in analogy to the dark matter has been dubbed the dark energy (DE). 
Although the DE phenomenon seems to be fairly well described by a simple cosmological constant $\Lambda $, its value has to be extremely fine-tuned in order to accurately reproduce observations. Furthermore, the attempt to relate the DE to the vacuum energy of quantum field theories fails in predicting the observed energy scale of DE by more than 100 orders of magnitude \cite{Weinberg_1988,Sahni_2002}.
Due to such exotic nature, dark matter and dark energy -- that
presently constitute about 95\% of the total energy density of the Universe -- represent one of the most intriguing phenomena of modern physics, and despite their very 
different observational manifestations several attempts have been made in order to investigate their possible mutual interactions \cite{Wetterich_1995,Amendola_2000,Farrar2004,Amendola_2004} or even to speculate about a possible common origin \cite{Padmanabhan_Choudhury_2002,Carturan_Finelli_2003,Bertacca_etal_2007} of these two dark components.
\ \\

In the present work we will explore the possibility that both the dark components of the Universe are actually more complex
than it is assumed by the standard cosmological $\Lambda $CDM model. On one side, we will allow the dark matter density to be
made up by more than one fundamental particle. On the other side, we will identify the dark energy with a dynamical light scalar field
and we will allow for direct interactions between the DE and the dark matter fluids, thereby accounting for a possible exchange
of energy-momentum in the dark sector of the Universe.
In particular, we will consider the possibility that Cold Dark Matter be composed by two different types of particles with opposite
interaction strength to the DE scalar field. This scenario represents a specific case of the more general models discussed in \cite{Brookfield_VanDeBruck_Hall_2008}, but differently from such general framework it requires the same number of parameters of a standard coupled DE cosmology
like the ones introduced by \cite{Wetterich_1995,Amendola_2000,Farrar2004}. Although clearly quite speculative, our proposed scenario
therefore represents a simple extension of widely studied models of the dark sector. The present work is then aimed at testing its
viability by assessing to which extent such kind of framework can be constrained using presently available observations. As we will show
in our discussion, a relevant portion of the parameter space of our model, significantly larger than what is presently allowed
for standard coupled DE scenarios, seems to be perfectly viable both at the background and at the linear perturbations level.

The paper is organized as follows. In Section~\ref{sec:model} we will define our class of models and introduce the main equations
and definitions that will be used in the rest of the analysis. In Section~\ref{sec:background} we will study the cosmological
background dynamics of the model for different choices of its parameters, and show to which extent the cosmic expansion history can be modified within our scenario. In Section~\ref{sec:linear} we will numerically study
the evolution of linear density perturbations and we will discuss the viability of the model based on the growth rate of large-scale structures.
Finally, in Section~\ref{sec:concl} we will draw our conclusions and we will suggest possible future developments in the investigation of our proposed scenario.

\section{Multiple Dark Matter and dark sector interactions}
\label{sec:model}

We consider a series of flat cosmological models including radiation, Cold Dark Matter (CDM) and Dark Energy (DE), 
where the role of the latter is played by a classical scalar field $\phi $ moving in a self-interaction potential $V(\phi )$, 
which is often referred to as the {\em Quintessence} \cite{Wetterich_1988,Ratra_Peebles_1988}. 
Without loss of generality for the aims of our analysis, in this work we will ignore the presence of baryonic matter
and we will assume an exponential form \cite{Lucchin_Matarrese_1984,Ferreira_Joyce_1998} for the scalar self-interaction potential $V(\phi )$:
\begin{equation}
V(\phi ) = Ae^{-\alpha \phi /M_{\rm Pl}}\,. 
\end{equation}

Following the initial proposal of \cite{Damour_Gibbons_Gundlach_1990,Wetterich_1995}, we allow for
an interaction within the dark sector of the Universe in the form of a direct exchange of energy-momentum between the DE
field $\phi $ and CDM particles. Such kind of interacting DE models have been widely studied in the literature concerning
their impact on the cosmic background evolution (see e.g. \cite{Amendola_2000,Farrar2004,Koyama_etal_2009,CalderaCabral_2009}), 
on the growth of linear density perturbations \cite{Amendola_2004,Pettorino_Baccigalupi_2008,Valiviita_etal_2009}, and also on the evolution of nonlinear structure formation 
\cite{Maccio_etal_2004,Mainini_Bonometto_2006,Baldi_etal_2010,Baldi_2011a,Li_Barrow_2011,CoDECS}.
All these studies have allowed to put constraints on the DE-CDM coupling constant
(see e.g. \cite{Bean_etal_2008,LaVacca_etal_2009,Xia_2009,Baldi_Viel_2010}) 
which is bound to a few percent of the gravitational interaction strength. In particular, coupling values of order unity 
(i.e. an interaction with the same strength as gravity) are ruled out based
on the strong impact that such interaction would have on the expansion history of the Universe as a consequence of
the meta-stable scaling solution between the DE and the CDM fluids during matter domination, which has been
dubbed the $\phi $-MDE phase ({\em $\phi $-Matter Dominated Epoch}, see \cite{Amendola_2000}) and which represents
one of the most characteristic features of standard coupled DE scenarios. Such $\phi $-MDE, with its associated
Early Dark Energy (EDE) component, determines a shift in the Matter-Radiation equality and a corresponding change
in the angular-diameter distance to last scattering that can be effectively constrained via CMB observations \cite{Bean_etal_2008,Xia_2009,Valiviita_etal_2009,Clemson_etal_2011}.

In order to evade such constraints, coupled DE scenarios with time-dependent couplings where the DE-CDM interaction 
strength is negligible at high redshifts and becomes significant only during the late stages of structure formation have been 
proposed in the literature (see e.g. \cite{Amendola_2004,Baldi_2011a}). These scenarios, however, require to define
{\em a priori} some specific form of the coupling evolution -- either in terms of the DE scalar field or in a more
phenomenological way as a function of the scale factor $a$ or the DE density -- suitable to
provide the desired suppression of the interaction at high redshifts. Although these variable-coupling models have proven
to easily evade background constraints still allowing significant effects of the DE-CDM coupling on structure formation
processes at low redshifts \cite{Baldi_2011a}, they require at least one additional free parameter with respect to standard coupled DE
scenarios with constant coupling.
\ \\

In the present work, we will move back to the case of constant couplings, and we will investigate a class of coupled DE cosmologies
for which coupling values of order unity and larger do not significantly affect the background evolution of the Universe. 
Differently from what has been assumed in most of previous studies, in fact, here we will consider the
possibility that the CDM fluid be composed by two different types of particles, with identical physical properties except for the sign
of their coupling to the DE scalar field $\phi $. 
Some other types of Multiple Dark Matter (MDM, hereafter) models have already been considered in the literature
in the context of Warm Dark Matter cosmologies (see e.g. \cite{Palazzo_etal_2007,Boyarsky_etal_2009,Maccio_etal_2012}) 
and also for the case of interacting DE scenarios (see e.g. \cite{Farrar2004,Huey_Wandelt_2006,Amendola_Baldi_Wetterich_2008,Brookfield_VanDeBruck_Hall_2008,Brax_etal_2010}).
In particular, {\em Brookfield, van de Bruck \& Hall 2008} (\cite{Brookfield_VanDeBruck_Hall_2008}, BVH08 hereafter) 
have considered a general setup where multiple matter fluids
interact with individual couplings with a classical DE scalar field, and highlighted for the first time some of the most
basic features of such MDM coupled DE scenario. Here we will focus on a specific case of the more general framework
defined in BVH08 by assuming that 
CDM is made of only two different particle species whose
individual couplings have the same absolute value but opposite signs.
In other words, in the present study we will consider the possibility that CDM particles carry
a {\em ``charge"} -- positive or negative -- with respect to their interaction with the DE field, and we will denote these two
distinct CDM species with the subscripts $+$ and $-$, respectively. This choice allows to restrict the parameter space
of the model with respect to the more general scenario of BVH08 and to reduce it to the same 
number of parameters of standard coupled DE with constant coupling, although still providing a much richer phenomenology. Furthermore, the fact that the coupling between DE and CDM is associated to a sort of {\em ``charge"}
of CDM particles might arise more naturally as a consequence of some new fundamental symmetry in the dark sector.

With such assumptions, the background evolution of the Universe will be described by the following system of dynamic field equations:
\begin{eqnarray}
\label{klein_gordon}
\ddot{\phi } + 3H\dot{\phi } + \frac{dV}{d\phi } &=& +C \rho _{+} - C \rho _{-}\,, \\
\label{continuity_plus}
\dot{\rho }_{+} + 3H\rho _{+} &=& -C \dot{\phi }\rho _{+} \,, \\
\label{continuity_minus}
\dot{\rho }_{-} + 3H\rho _{-} &=& +C \dot{\phi }\rho _{-} \,, \\
\label{continuity_radiation}
\dot{\rho }_{r} + 4H\rho _{r} &=& 0\,, \\
\label{friedmann}
3H^{2} &=& \frac{1}{M_{{\rm Pl}}^{2}}\left( \rho _{r} + \rho _{+} + \rho _{-} + \rho _{\phi} \right)\,,
\end{eqnarray}
where an overdot represents a derivative with respect to the cosmic time $t$, $H\equiv \dot{a}/a$ is the Hubble function,
the CDM density is given by $\rho _{\rm CDM} = \rho _{+}+\rho _{-}$, and $M_{\rm Pl}\equiv 1/\sqrt{8\pi G}$ is the reduced Planck mass with $G$ the Newton's constant.
The dimensional coupling constant $C$ is defined as:
\begin{equation}
C\equiv \sqrt{\frac{2}{3}}\frac{1}{M_{\rm Pl}}\beta \,,
\end{equation}
with $\beta = {\rm const.} \geq 0$ being the standard definition (see e.g. \cite{Amendola_2004,Baldi_2011a,CoDECS}) of the
dimensionless coupling that sets the strength of the interaction between DE and CDM.

One of the most basic features of such interaction (see e.g. \cite{Amendola_2000,Baldi_etal_2010,Baldi_2011b}) is the variation of the mass of CDM particles as a consequence of the
dynamical evolution of the DE scalar field, according to the equation:
\begin{equation}
\label{mass_variation}
\frac{d\ln \left[ M_{\pm}/M_{\rm Pl}\right] }{dt} = \mp C\dot{\phi }\,,
\end{equation}
where $M_{\pm }$ is the mass of a CDM particle of the positively ($+$) or negatively ($-$) coupled species, and
where we have now taken into account the opposite variation of the mass of particles of the two different CDM types associated to their opposite couplings.
Due to this different mass evolution, the relative abundance of the two CDM species does
also vary in time whenever $\dot{\phi } \neq 0$, giving rise to a time-dependent asymmetry between the two CDM fluids. To quantify this concept, we introduce
the dimensionless asymmetry parameter $\mu $, defined as:
\begin{equation}
\label{eta}
\mu  \equiv \frac{\Omega _{+} - \Omega _{-}}{\Omega _{+} + \Omega _{-}}\,,
\end{equation}
where the fractional density parameters $\Omega _{i}$ are defined in the usual way as:
\begin{equation}
\Omega _{i} \equiv \frac{\rho _{i}}{3H^{2}M_{\rm Pl}^{2}}\,.
\end{equation}
As already pointed out by BVH08, the background dynamics of a general MDM coupled DE model with constant couplings $\beta _{j}$
is equivalent to that of a coupled DE scenario with a single CDM fluid and with a time-dependent coupling
$\beta _{\rm eff}$ given by:
\begin{equation}
\beta _{\rm eff} \equiv \frac{\sum_{j}\beta _{j}\Omega _{j}}{\sum_{j}\Omega _{j}} \,,
\end{equation}
which for our specific model then simply reads:
\begin{equation}
\label{effective_coupling}
\beta _{\rm eff }= \beta \left( \frac{\Omega _{+}}{\Omega _{\rm CDM}} - \frac{\Omega _{-}}{\Omega _{\rm CDM}}\right) = \beta \mu  \,.
\end{equation}
Such equivalence provides a self-regulating mechanism for dark sector interactions since the global effective coupling is dynamically suppressed
during matter domination, as we will discuss in the next Section.
According to Eqs.~(\ref{eta}) and (\ref{effective_coupling}), a standard coupled DE scenario -- i.e. a model with only one CDM fluid interacting with a positive
coupling with the DE scalar field -- would correspond to a value of $\mu  = +1$ during the whole expansion history of the Universe, while
$\mu  = -1$ would also represent a single-CDM model but with negative constant coupling. 
Finally, from Eq.~(\ref{effective_coupling}) one can see
that the effective dimensionless coupling acting on the scalar field $\phi $ identically vanishes for $\mu  = 0$, such that for a perfectly symmetric state with $\Omega_{+} = \Omega _{-}$ the 
DE field does not experience any coupling to CDM, thereby behaving at the background level like a minimally coupled {\em Quintessence} field.

\section{Background evolution}
\label{sec:background}

In order to study the cosmological evolution of the MDM coupled DE models defined above, we numerically integrate
the system of dynamical background equations (\ref{klein_gordon}-\ref{friedmann}) for different values of the
dimensionless coupling $\beta $ and the primordial asymmetry parameter $\mu  $, which we denote with $\mu  _{\infty }$.
It is important to notice here that while the coupling $\beta $ is an intrinsic parameter of the model, which defines the interaction
strength between DE and CDM particles, the primordial asymmetry $\mu  _{\infty }$ is simply a way to parametrize the
initial conditions of the system by fixing the relative abundance of the two CDM species in the early Universe, just as the 
baryon-to-photon ratio $\eta $ sets the value of the primordial ratio $\Omega _{b}/\Omega _{r}$. 
Therefore, our proposed MDM coupled DE scenario has only
two intrinsic parameters, namely the slope of the self-interaction potential $\alpha $ and the coupling strength $\beta $, i.e.
the same number of parameters of a standard coupled DE model with constant coupling. In this respect, our specific MDM coupled DE models represent a particularly appealing realization of the general scenario of BVH08 which requires the smallest possible number of free parameters.

As a reference cosmology we consider a minimally coupled scalar field scenario with $\beta = 0$ and with a slope of the 
DE self-interaction potential $\alpha = 0.08$, and we integrate the corresponding system of equations (\ref{klein_gordon}-\ref{friedmann})
backwards in time until deep into the radiation dominated epoch (i.e. until $\Omega _{r} > 0.999$) following the integration procedure 
discussed in \cite{Baldi_2011a}. We start the integration at $z=0$ with a set of cosmological parameters compatible with the latest
CMB constraints from WMAP7 \cite{wmap7}.
For such reference scenario, which we denote EXP000, due to the absence of coupling the asymmetry parameter
$\mu  $ is completely irrelevant since the energy density of both CDM species scales like $a^{-3}$ and any chosen value of $\mu $
in the allowed range $[-1,+1]$ will therefore remain constant in time without affecting the expansion history of the Universe, which
we verified to be indistinguishable from a $\Lambda $CDM cosmology with the same cosmological parameters.
%%%%%%%%%%%%%%%%%%%%
\begin{table}
\caption{A selected sub-sample of all the MDM coupled DE models investigated in the present work. The upper half of the table displays some properties of symmetric models ($\mu _{\infty } = 0$) with different values of the coupling $\beta $, while the lower half of the table shows the same properties for a series of models with fixed coupling $\beta = 2$ and different values of the primordial asymmetry parameter $\mu _{\infty }$. As the table shows, symmetric models do not appreciably affect the background evolution of the Universe even for very large coupling values.}
\label{tab:models}
\begin{tabular}{@{}lccc | cccccc@{}}
\hline
Model & $\alpha $ & $\beta$ & $\mu  _{\infty }$ & $\Omega _{\phi }$ & $\Omega _{\rm CDM}$ & $\mu _{0}$ & $w _{\phi }$ & $z_{\rm eq}$ & $z_{\rm de}$ \\
\hline
EXP000 & $0.08$ & $0.0$ & -- & $0.729$ & $0.271$ & $0.0$ & $-0.999$ & $3164.29$ & $0.377$\\
\hline
EXP010+0.0 & $0.08$ & $0.5$ & $0.0$ &  $0.729$ & $0.271$ & $-0.01$ & $-0.999$ & $3164.29$ & $0.391$\\
EXP020+0.0 & $0.08$ & $1.0$ & $0.0$ &  $0.729$ & $0.271$ & $-0.02$ & $-0.999$ & $3164.29$ & $0.391$\\
EXP030+0.0 & $0.08$ & $1.5$ & $0.0$ & $0.729$ & $0.271$ & $-0.03$ & $-0.999$ & $3164.29$ & $0.391$\\
EXP040+0.0 & $0.08$ & $2.0$ & $0.0$ & $0.729$ & $0.271$ & $-0.03$ & $-0.999$ & $3164.29$ & $0.391$\\
EXP200+0.0 & $0.08$ & $10.0$ & $0.0$ & $0.729$ & $0.271$ & $-0.02$ & $-0.999$ & $3164.29$ & $0.391$\\
\hline
EXP040+0.1 & $0.08$ & $2.0$ & $+0.1$ & $0.729$ & $0.271$ & $-0.03$ & $-0.999$ & $3164.29$ & $0.391$\\
EXP040-0.1 & $0.08$ & $2.0$ & $-0.1$ & $0.731$ & $0.269$ & $-0.03$ & $-0.999$ & $3164.29$ & $0.391$\\
EXP040+0.5 & $0.08$ & $2.0$ & $+0.5$ & $0.751$ & $0.248$ & $-0.03$ & $-0.999$ & $2750.77$ & $0.433$\\
EXP040-0.5 & $0.08$ & $2.0$ & $-0.5$ & $0.762$ & $0.238$ & $-0.03$ & $-0.999$ & $2750.77$ & $0.462$\\
EXP040+0.9 & $0.08$ & $2.0$ & $+0.9$ & $0.851$ & $0.149$ & $-0.05$ &$-0.999$ & $1393.09$ & $0.786$\\
EXP040-0.9 & $0.08$ & $2.0$ & $-0.9$ & $0.869$ & $0.131$ & $-0.05$ & $-0.999$ & $1393.09$ & $0.878$\\
\hline
\end{tabular}
\end{table}
%%%%%%%%%%%%%%%%%%%%

After integrating backwards in time the background dynamic equations of our reference model EXP000, we assume the final state of this
integration at very high redshift ($z_{i}\sim 10^{7}$) as initial conditions for the forward integration of all the other MDM coupled DE models 
considered in the present work. In particular, we will
always assume the final values of $\phi $ and $\dot{\phi }$ of our reference backwards integration as the scalar field initial
conditions for all the different models under consideration. 
This two-way integration strategy ensures to have a reference model -- indistinguishable from $\Lambda $CDM -- with exactly the desired cosmological parameters at $z=0$, and 
to avoid possible instabilities in the backwards integration of the coupled DE scenarios.
Clearly, such procedure will not necessarily provide
a viable cosmological evolution for all the models under investigation, but this is not the goal of the present work
which rather aims at exploring the main features of MDM coupled DE models by quantifying their deviations from a reference
standard cosmological scenario with the same initial conditions. We performed such integration for a large number of different
models by varying the value of the coupling $\beta $ and of the primordial asymmetry $\mu _{\infty }$, and we
summarize a selected sub-sample of such models -- with their parameters and some results of the background integration -- in Table
~\ref{tab:models}, which includes a few symmetric models (i.e. with $\mu _{\infty } = 0$) with different coupling values ($\beta =0.5-10.0$)
as well as some asymmetric models (i.e. with $\mu _{\infty } \ne 0$) only for one specific value of the coupling, $\beta = 2$.
Without loss of generality we assumed for all the models the same slope $\alpha $ of the scalar self-interaction potential since
varying $\alpha $ does not show any effect on the main features of the MDM coupled DE models under discussion.

Relying on our large sample of integrated background cosmologies for MDM coupled DE models, in the next subsections we will show -- broadly confirming the previous findings of BVH08 -- 
that the presence of two dark matter species with opposite couplings to the DE scalar field
can significantly loosen the present background constraints on the CDM-DE coupling $\beta $.
Furthermore, extending previous analyses, we will also investigate
how such screening effect depends on the parameters and on the initial conditions of the model, $\beta $ and $\mu _{\infty }$.
Our results, besides confirming previous outcomes on MDM coupled DE models, will therefore also explore the stability of the scenario
with respect to a possible primordial asymmetry between the two CDM components, and will significantly extend the range of parameters for which similar models have been previously tested. We want to stress once more at this point that choosing the specific case of opposite couplings with the same absolute value for the two different CDM species does not only allow to reduce the parameter space of the model to the same number of parameters as standard coupled DE, but also provides a direct connection to a possible origin of the DE-CDM coupling as the manifestation of some new fundamental symmetry characterizing the dark sector.

\subsection{Symmetric models}

We start our analysis from initially symmetric states, i.e. from models that start in the early Universe with an even abundance
of the two CDM species, such that $\Omega _{+}(z_{i}) = \Omega _{-}(z_{i})$, or in other terms $\mu  _{\infty } = 0$.
Such condition is consistent with the idea of the two CDM particles being degenerate in mass and formed out of thermal equilibrium
processes in the early Universe. However, this symmetry is bound to be rapidly broken due to the 
dynamical evolution of the scalar field, i.e. the fact that at high redshifts $\dot{\phi } \neq 0$, which will determine a different
scaling for the two CDM species according to Eq.~(\ref{mass_variation}). 

Starting from the same initial conditions as the reference model EXP000 at $z_{i}$, we integrate the system of equations (\ref{klein_gordon}-\ref{friedmann})
to the present time for five different values of the coupling $\beta $, namely $\beta = 0.5\,,1\,,1.5\,,2\,,10$. Even the lowest of these coupling values ($\beta = 0.5$)
is ruled out at more than 6$\sigma $ for standard coupled DE scenarios, for which it
would determine a cosmological evolution starkly incompatible with CMB, Large Scale Structure, and Lyman-$\alpha $ observations \cite{Bean_etal_2008,Xia_2009,Baldi_Viel_2010}.
On the other hand, a coupling as large as $\beta = 10$ in a standard coupled DE scenario would even prevent the existence of a 
Matter Dominated Epoch (see \cite{Amendola_2000})
and would feature a direct switch from radiation domination to an accelerated DE dominated regime, thereby determining a completely unrealistic cosmology.

\begin{figure}
\begin{minipage}{70mm}
\includegraphics[width=\linewidth,height=57mm]{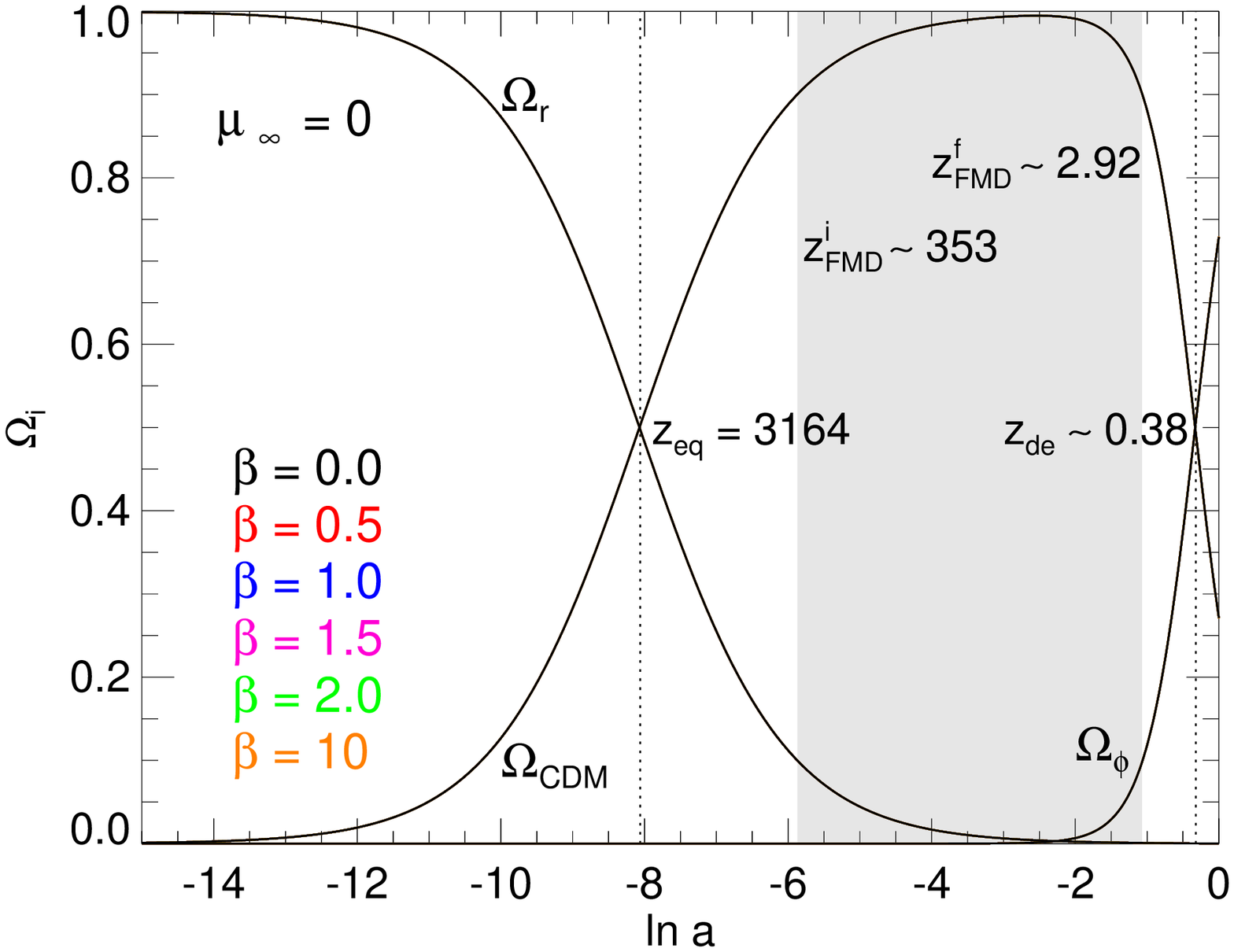}
\caption{The background evolution of an uncoupled cosmology identical to $\Lambda $CDM (black curves) and of a series of MDM coupled DE models with symmetric initial conditions and different coupling values. The grey-shaded area represents the Full Matter Dominated phase ($\Omega _{\rm CDM}>0.9$) and the vertical dotted lines correspond to CDM-radiation and CDM-DE equivalence. As the plot shows, symmetric MDM coupled DE models with couplings as large as $\beta = 10$ are completely indistinguishable from $\Lambda $CDM in the background.}
\label{fig:background_symm}
\end{minipage}
\hfil
\begin{minipage}{70mm}
\includegraphics[width=\linewidth,height=57mm]{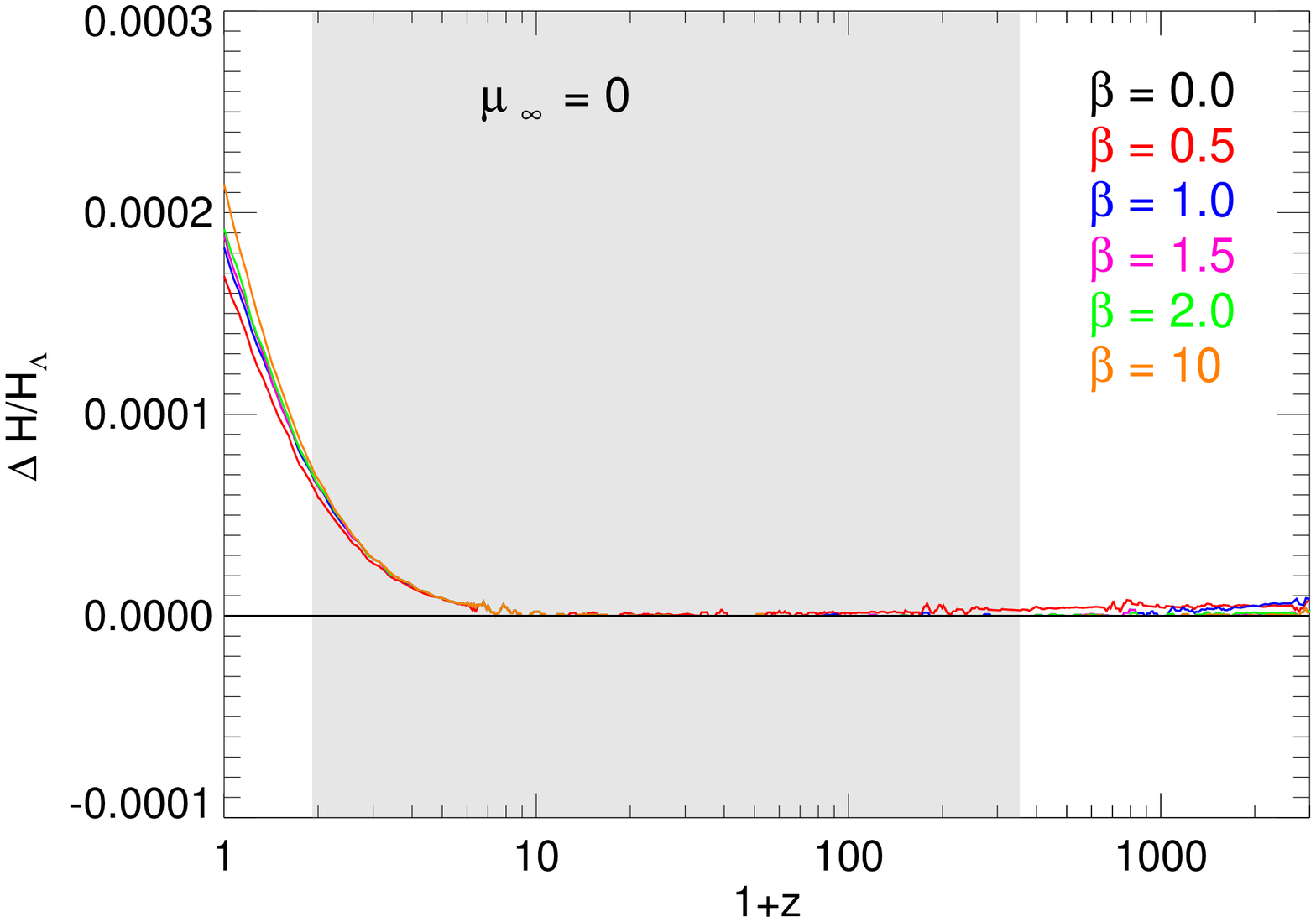}
\caption{The ratio of the Hubble function $H(z)$ over the $\Lambda $CDM Hubble function $H_{\Lambda }$ for all the symmetric MDM coupled DE models of Fig.~\ref{fig:background_symm}. The expansion histories of all the models are indistinguishable from $\Lambda $CDM until the end of matter domination, where some deviations start to appear. However, such deviations never exceed a few hundredths of a percent and are therefore clearly undetectable. Therefore, the expansion histories of symmetric MDM coupled DE models are completely indistinguishable from $\Lambda $CDM even for couplings as large as $\beta = 10$.}
\label{fig:Hratio_symm}
\end{minipage}
\end{figure}

In the context of MDM coupled DE, instead, the effect of such large couplings on the expansion history of the Universe is
suppressed by the balance between the opposite interactions of the two CDM species, which determines a very mild impact
of the coupling on the cosmological background evolution. Figure~\ref{fig:background_symm} shows the cosmological evolution
of the fractional density of radiation, CDM, and DE, as a function of the e-folding time defined as the natural logarithm of the scale
factor $a$, for the reference model EXP000 (black curves) and for all the symmetric MDM coupled DE models under consideration 
in the present work (colored curves). The vertical dotted lines correspond to the CDM-radiation equivalence and to the CDM-DE
equivalence, that take place in the reference model at $z_{\rm eq}\sim 3164$ and $z_{\rm de}\sim 0.38$, respectively, while
the grey-shaded area indicates the epoch when $\Omega _{\rm CDM} > 0.9$, which we denote as {\em ``Full Matter Dominated"} (FMD) phase.
In most of the remaining figures of this work we will highlight with grey shading -- whenever relevant -- the range of extension of
the FMD phase corresponding to the reference model EXP000.

By having a look at Fig.~\ref{fig:background_symm} it is immediately clear that none of the symmetric MDM coupled DE models has a significant impact on
the background evolution of the Universe, since the corresponding colored curves cannot be distinguished from the reference model and are
actually completely hidden behind the black curves representing the uncoupled case, which in turn is undistinguishable from the
standard concordance $\Lambda $CDM cosmology. Therefore, Fig.~\ref{fig:background_symm} shows that all the MDM coupled DE
models with symmetric initial conditions, even for a coupling as large as $\beta = 10$, are completely indistinguishable from $\Lambda $CDM in the background.

The impact of the different scenarios on the expansion history is better quantified by Fig.~\ref{fig:Hratio_symm},
where we plot the ratio of the Hubble function of each model as computed with our numerical integrations, over the Hubble function
of the reference uncoupled model, which we denote with $H_{\Lambda }$ being it indistinguishable from the Hubble function
of the concordance $\Lambda $CDM scenario. The figure clearly shows that for all models $H(z)$ is also indistinguishable from 
$H_{\Lambda }$ from $z_{\rm eq}$ down to $z\sim 5$. Some deviation starts to develop at later times, in correspondence with
the end of the FMD epoch and the onset of DE domination; such deviations however never exceed the level of a few hundredth
of a percent, even for the most extreme scenario with $\beta =10$, and are therefore completely irrelevant from an observational point of view. 
This plot therefore confirms that
for symmetric initial conditions MDM coupled DE models do not appreciably affect the background expansion history of the Universe 
even for very large values of the coupling constant $\beta $. It is nevertheless interesting to notice already at this stage that
the only small deviations from the reference uncoupled scenario appear in correspondence with 
the emergence of a DE component in the Universe, while until CDM dominates the cosmic energy budget, the system is kept 
on an effectively uncoupled trajectory. 

\begin{figure}[t]
\sidecaption
\includegraphics[scale=0.4]{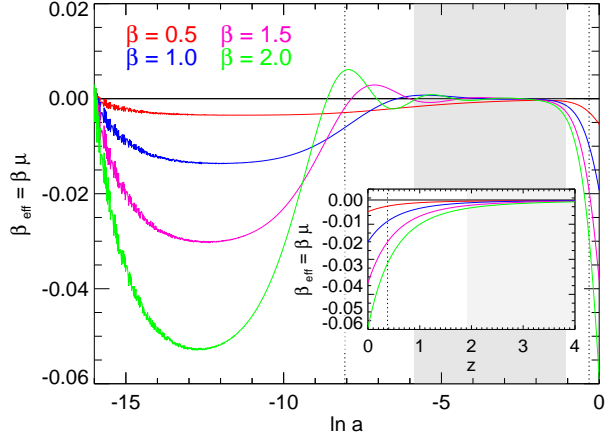}%
\caption{The effective coupling of symmetric MDM coupled DE models as a function of the e-folding time. The system is attracted towards the uncoupled critical point $\mu = 0$ during matter domination, despite the deviation from the initial symmetry that develops in radiation domination. After the end of matter domination these cosmologies evolve again to effectively negatively coupled systems due to the onset of DE domination. The small box shows a zoom of the same quantities as a function of redshift for $z<4$, and the grey-shaded regions indicate the FMD phase.}
\label{fig:beff_z_symm}
\end{figure}

The fact that a symmetric MDM coupled DE scenario evolves -- as we just showed -- like an uncoupled system might look an obvious
consequence of Eq.~(\ref{effective_coupling}), which shows that for a symmetric state $\mu = 0$ the effective coupling $\beta _{\rm eff}$ acting on the DE
scalar field identically vanishes. However, the situation is not so simple and the result that we just discussed is in the end not so obvious. In fact, as we already mentioned above, 
the initial symmetry of the system that we enforce by setting $\mu _{\infty } = 0$ is bound to be rapidly broken by the dynamical evolution of the scalar field $\phi $
that starts at high redshifts with a positive velocity $\dot{\phi }_{i} > 0$. Therefore, even if starting from a symmetric situation, an asymmetry between
the two CDM species will soon develop according to Eq.~(\ref{mass_variation}). The situation is depicted clearly in Fig.~\ref{fig:beff_z_symm}
where the evolution of the effective coupling $\beta _{\rm eff}$ as a function of the e-folding time is displayed for couplings as large as $\beta = 2$. All the models start with $\beta _{\rm eff} = 0$ but soon develop a non-zero effective coupling during radiation domination. However, when approaching the matter-radiation equivalence the system
is dragged again towards $\beta _{\rm eff} = 0$ and during the FMD phase the effective coupling remains close to zero, while its absolute value starts to grow again only at the end of FMD when DE takes over.

Such behavior is determined by a new matter-dominated critical point in the phase-space of the background dynamical
system (\ref{klein_gordon}-\ref{friedmann}) -- that was first identified by BVH08 -- defined by the condition:
\begin{equation}
\sum_{j} \beta _{j}\Omega _{j} = 0
\end{equation}
with $j$ ranging over all the coupled matter species. For the specific case of our MDM coupled DE models (i.e. $j=\left\{ +\,,-\right\} $)
this new critical point simply turns into the condition $\mu = 0$, to which the system is therefore attracted during matter domination.

In particular, the small box in Fig.~\ref{fig:beff_z_symm} shows a zoom on the evolution of the effective coupling as a function of redshift
for $z<4$. As one can see from the plot, the value of the effective coupling starts to deviate from zero in correspondence to the end of the FMD phase, and steeply evolves to progressively more negative values towards $z=0$. This corresponds to the fact that the symmetry
between the two CDM fluids that holds in matter domination is broken again at late times in favor of the negatively coupled species whose
particles mass starts growing in time, while the positively coupled species features the opposite trend.
This is a consequence of the dynamical evolution of the scalar field that after being frozen during matter domination in the minimum of the effective potential defined by:
\begin{equation}
\frac{d V_{\rm eff}}{d\phi }\equiv \frac{dV}{d\phi} -C\rho _{+} + C\rho _{-}\,,
\end{equation}
starts moving again as soon as DE takes the lead of the cosmic budget. MDM coupled DE models therefore naturally provide a time-dependent effective coupling of the form that was proposed in \cite{Baldi_2011a} without imposing {\em a priori} any specific form for the coupling evolution.

\subsection{Asymmetric models}

We move now to explore the possibility that the two CDM species do not share the same relative abundance in the early Universe.
Such asymmetric initial state might arise if one or both the CDM components are created by non-thermal processes or if some early
dynamics of the DE scalar field (as e.g. a kination phase during radiation domination) pushes the system significantly off from the
symmetric state $\mu _{\infty } = 0$. We will therefore integrate again forward in time the system of background equations (\ref{klein_gordon}-\ref{friedmann}) starting from the same initial conditions of the reference scenario EXP000, but this time 
for every value of the coupling $\beta $ (except the most extreme $\beta = 10$) we will vary the initial asymmetry parameter considering the values $\mu _{\infty } = \pm 0.1\,, \pm 0.5\,, \pm 0.9$.

This procedure will clearly not necessarily provide viable background evolutions, since the mutual screening of the DE-CDM
coupling is weakened by the asymmetry between the two CDM species, such that the initial effective coupling is correspondingly large. However,  
our aim here is mainly
to investigate to which extent an early asymmetry can affect the subsequent cosmological evolution of a MDM coupled DE model 
and which level of primordial asymmetry might still provide a viable expansion history for a given coupling value.

\begin{figure}[h]
\begin{minipage}{70mm}
\includegraphics[width=\linewidth,height=57mm]{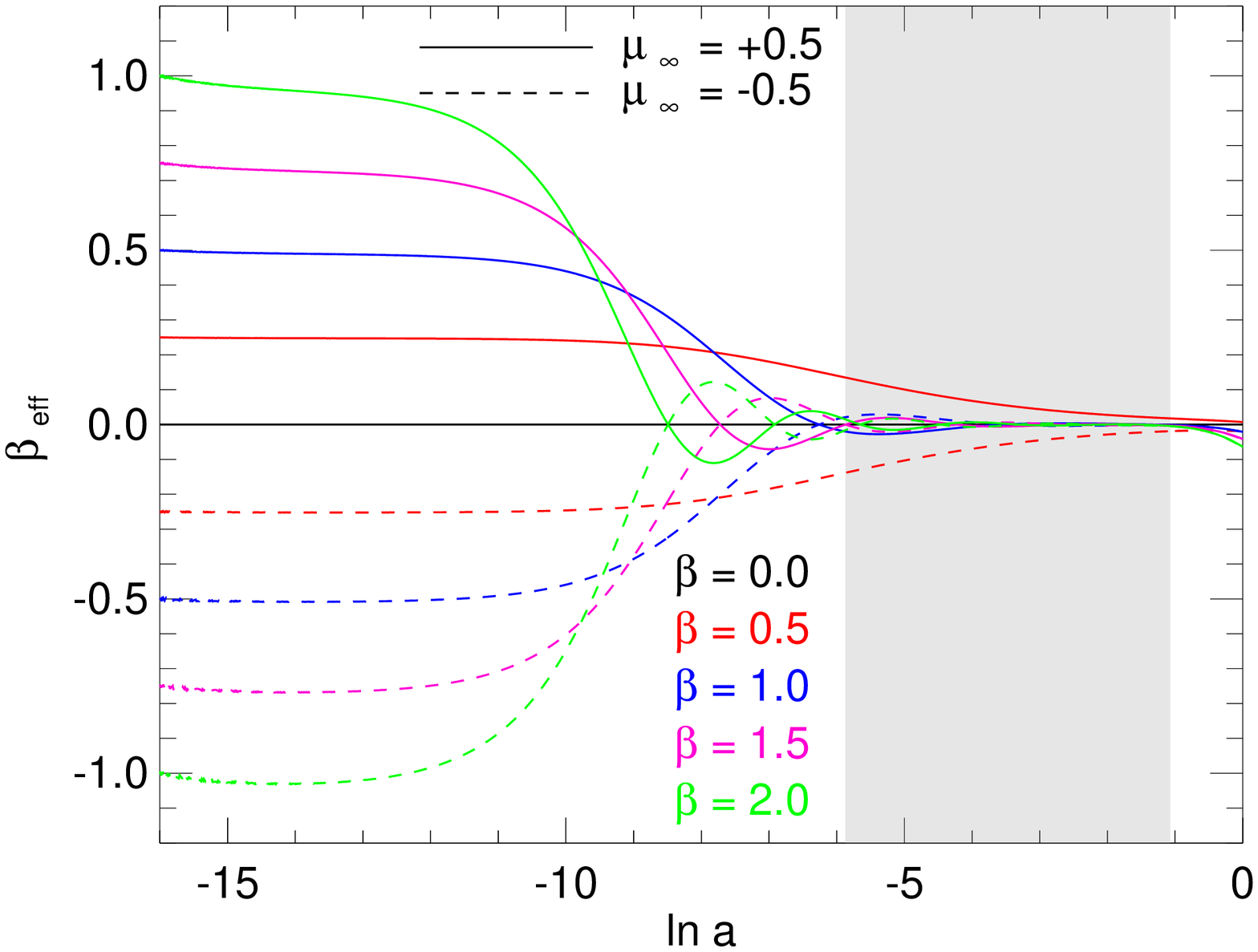}
\caption{The effective coupling for a series of asymmetric MDM coupeld DE models ($\mu _{\infty }\neq 0$) with different coupling values. All the models, irrespectively of the initial asymmetry, are attracted towards a symmetric state during matter domination, but the dragging appears more efficient for models with a larger coupling $\beta $.}
\label{fig:beff_asymm}
\end{minipage}
\hfil
\begin{minipage}{70mm}
\includegraphics[width=\linewidth,height=57mm]{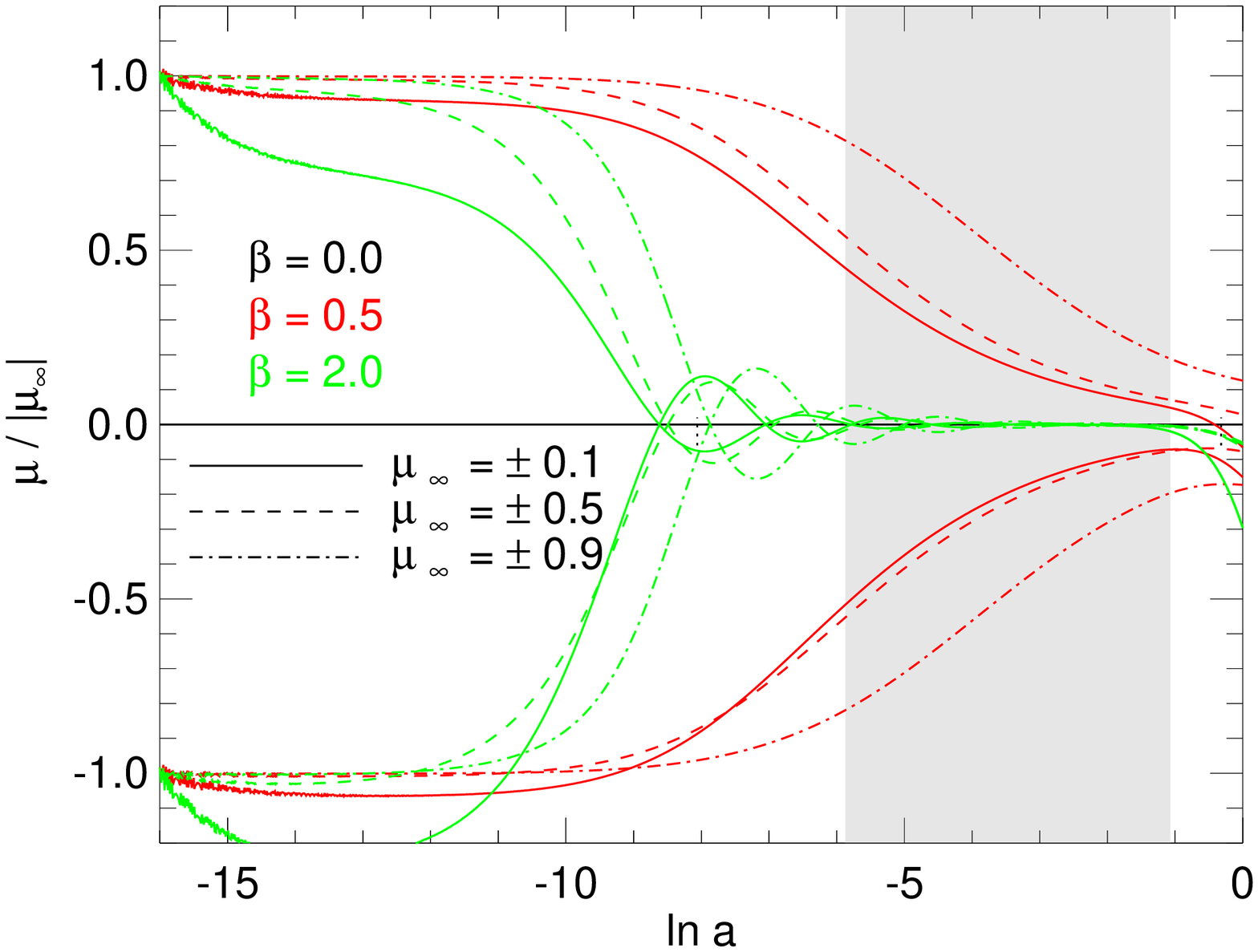}
\caption{The relative suppression of the primordial asymmetry for MDM coupled DE models with different values of $\mu _{\infty }$ and for two extreme values of the coupling, $\beta = 0.5$ (red) and $\beta = 2$ (green). The larger coupling proves to be much more effective in dragging the system to the symmetric critical point in matter domination.}
\label{fig:muratio_asymm}
\end{minipage}
\end{figure}

In Figure~\ref{fig:beff_asymm} we show the evolution of the effective coupling $\beta _{\rm eff}$ as a function
of the e-folding time for different values of the coupling $\beta $ and for a specific amount of primordial
asymmetry between the two CDM species, namely $|\mu _{\infty }| = 0.5$. As the plot shows, although the
primordial effective coupling can be large due to the reduced screening between the two matter fluids, it 
steeply decays towards the end of radiation domination and during the FMD epoch -- indicated by the gray-shaded area --
it remains close to zero. This shows that the primordial asymmetry is progressively washed out by the dynamical evolution
of the universe as the system is attracted towards the uncoupled critical point $\mu = 0$ in matter domination. However,
it is particularly interesting to notice how the efficiency with which the primordial asymmetry is diluted is inversely proportional
to the coupling strength $\beta $, with the weakest coupling model ($\beta =0.5$, red curves) showing in the FMD phase a larger effective
coupling than the model with the strongest coupling ($\beta = 2$, green curves) for the same value of the primordial
asymmetry $\mu _{\infty }$. In other words, if any asymmetry between the two CDM species is present in the early Universe, 
a larger coupling $\beta $ would more effectvely and rapidly suppress it and then result in a smaller effective coupling $\beta _{\rm eff}$ during matter domination, which would then
determine a weaker impact on the cosmic expansion history at late times. This somewhat counterintuitive result is confirmed and better
described by Figure~\ref{fig:muratio_asymm}, where the evolution of the asymmetry parameter $\mu $ normalized to its primordial
amplitude $|\mu _{\infty }|$ is displayed for several values of $\mu _{\infty }$ and for the two extreme values of the coupling $\beta =0.5$ and
$\beta = 2$. The figure clearly shows that larger initial asymmetries take longer to be dragged to the uncoupled critical point $\mu =0$,
but also that for the same primordial asymmetry $\mu _{\infty }$ a larger coupling determines a faster and more efficient suppression of the asymmetry during matter domination. Following the evolution of the most weakly coupled model ($\beta = 0.5$, red curves) one can
see that even for small primordial asymmetries ($|\mu _{\infty }|=0.1$, solid lines) the system does not fully reach the symmetric critical point during matter domination, while
this happens well before the end of radiation domination in the most strongly coupled case ($\beta = 2$, green curves) for any value of $|\mu _{\infty }|$.

These results reinforce the conclusion that MDM provides a self-regulating mechanism for dark sector interactions as the effective
background coupling is dragged towards zero in matter domination independently on the primordial relative abundance of the two CDM
fluids. Furthermore, such self-regulating mechanism is more efficient for larger values of the coupling constant which provide a 
faster and more effective screening of the interaction for the evolution of the cosmological background.

\begin{figure}[t]
{\bf (a)}~
\includegraphics[scale=0.35]{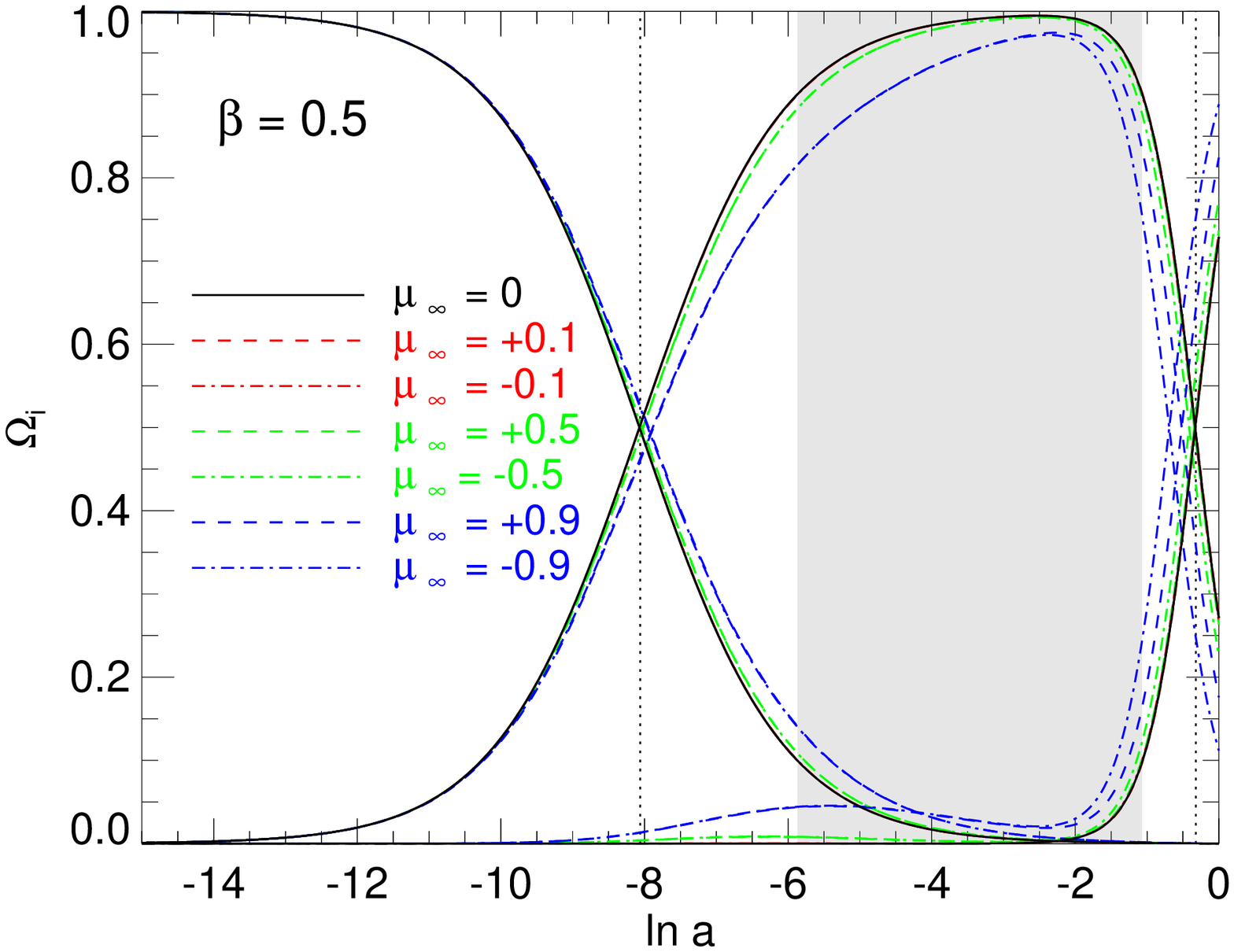}
\hfil
\includegraphics[scale=0.35]{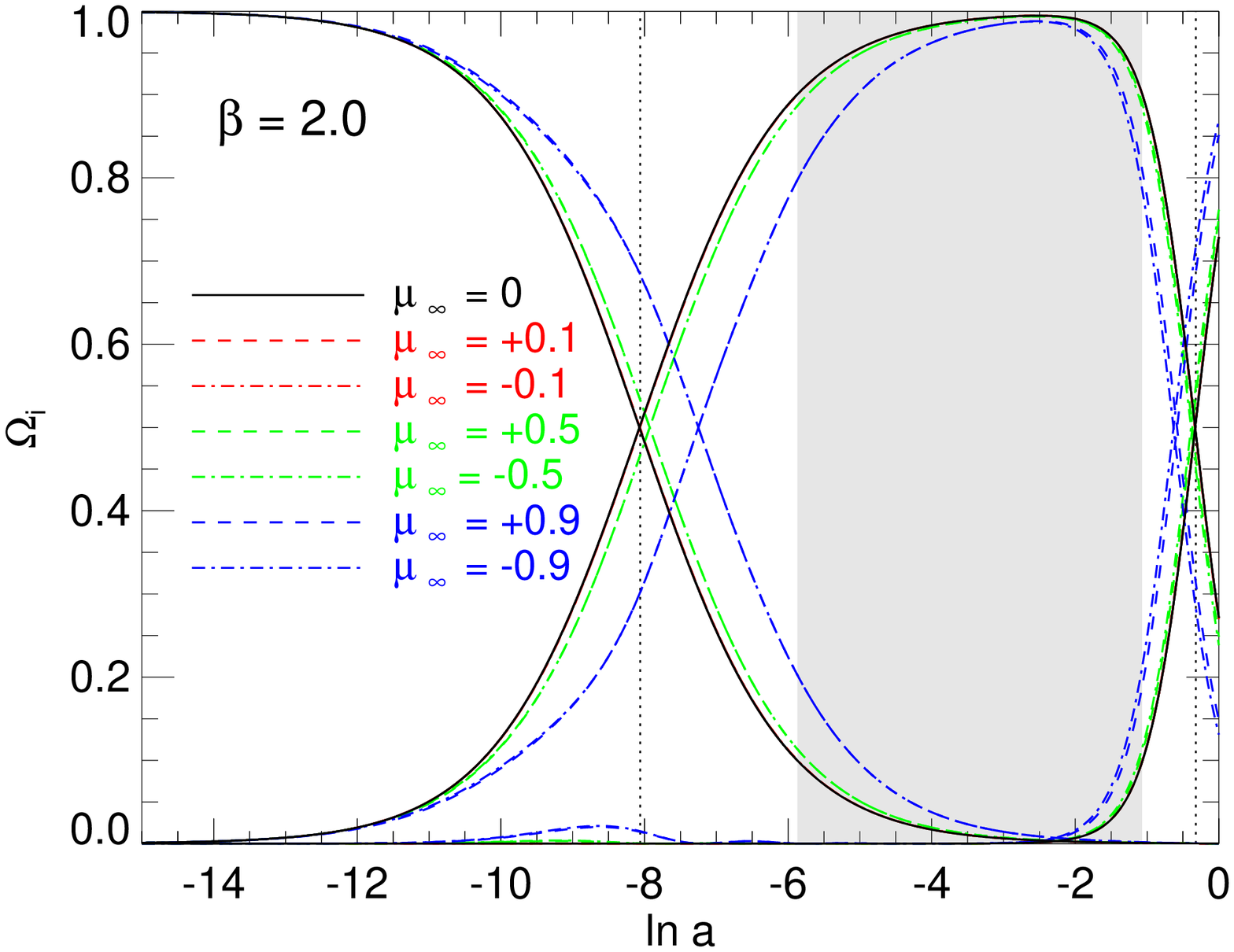}~{\bf (b)}
\caption{The background evolution for two asymmetric MDM coupled DE models with coupling $\beta = 0.5$ ({\bf a}) and $\beta = 2$ ({\bf b}) for different values of the primordial asymmetry $\mu _{\infty }$. The impact on the background dynamics is large for very large primordial asymmetries ($\mu _{\infty }=\pm 0.9$, blue curves) but for asymmetries of the order of $|\mu _{\infty }| = 0.1$ the background evolution is still practically indistinguishable from $\Lambda $CDM.}
\label{fig:background_asymm}
\end{figure}

To give an idea of the global impact of a primordial asymmetry between the two CDM species on the cosmic expansion history,
we plot in Fig.~\ref{fig:background_asymm} the background evolution for different values of the primordial asymmetry $\mu _{\infty}$ 
and for two values of the coupling, namely $\beta =0.5$ (\textbf{a}) and $\beta = 2$ (\textbf{b}). In both cases, although clearly in a more 
prominent way for the larger coupling, a primordial asymmetry as large as $|\mu _{\infty }| = 0.9$ (blue) determines a significant modification of the 
background evolution which is  clearly sufficient to rule out the models, while for $|\mu _{\infty }| = 0.1$ (red) the curves are completely indistinguishable from the reference scenario. The intermediate values $|\mu _{\infty }| = 0.5$ (green) would require a more quantitative comparison with actual data, as
their viability might be easily excluded for the stronger coupling but possibly not for the weaker one. In any case, these plots show that even for large
values of the coupling a 10\% asymmetry in the early Universe between the two matter species would not appreciably spoil the background expansion history. Therefore, the conclusion that MDM provides a self-regulating mechanism for dark interactions does not necessarily require any fine-tuning of the 
initial relative abundance of the two CDM species, which clearly makes the case for such models more natural and the whole argument more robust.

\begin{figure}
\sidecaption
\includegraphics[scale=0.4]{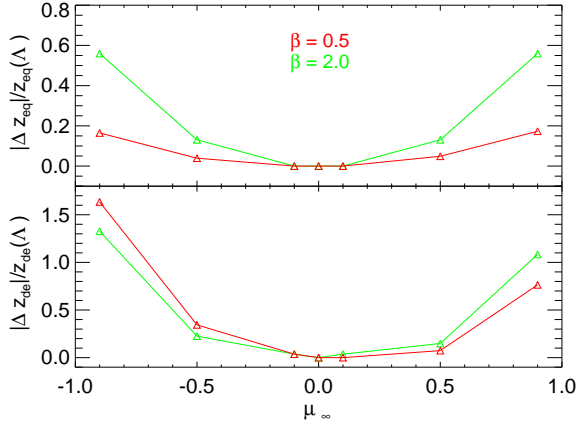}
\caption{The relative shift in the matter-radiation equivalence redshift $z_{\rm eq}$ (upper panel) and in the matter-DE equivalence redshift $z_{\rm de}$ (lower panel) for MDM coupled DE models  with couplings $\beta = 0.5$ (red) and $\beta = 2$ (green) as a function of the primordial asymmetry parameter $\mu _{\infty }$. As the plot shows, while at high redshifts the impact of positive and negative primordial asymmetries is indistinguishable, at low redhshifts this is no longer true and negative asymmetries have in general a larger impact than positive ones.}
\label{fig:equivalence_shift}
\end{figure}

Another interesting feature emerging from Fig.~\ref{fig:background_asymm} concerns the different impact of opposite values of the primordial
asymmetry. In fact, if at very high redshifts (i.e. before the beginning of the FMD epoch) positive and negative asymmetries show identical
evolutions, this is no longer true at later times, when a clear difference between these two cases is visible in the figures. Furthermore, such 
different evolution between positive and negative primordial asymmetries appears to be more pronounced for weaker values of the coupling $\beta $.
This is what is more quantitatively described in Figure~\ref{fig:equivalence_shift} where we plot the relative shift of the matter-radiation equivalence redshift $z_{\rm eq}$
(upper panel) and of the matter-DE equivalence redshift $z_{\rm de}$ (lower panel) as a function of the primordial asymmetry $\mu _{\infty }$ for two different values of the coupling $\beta $. As the figure shows, while the impact of positive and negative asymmetries is indistinguishable at the
redshift of matter-radiation equivalence, with larger couplings giving rise to larger shifts, the same is no longer true at low redshifts, where negative
asymmetries have a larger impact than positive ones for any given coupling, and where also the hierarchy of couplings is inverted for negative asymmetries, with weaker couplings determining a larger shift of the matter-DE equivalence time. This is due to the lower efficiency of  weak couplings in suppressing the primordial asymmetry during matter domination, as clearly shown in Figure~\ref{fig:muratio_asymm}. 

All these features of the background dynamics of MDM coupled DE models give an idea of the rich phenomenology that can arise
from the simple assumption that CDM particles are {\em ``charged"} with respect to their interactions with a DE scalar field, even without
significantly affecting the overall expansion history of the Universe. In the next Section we will show how this self-regulating mechanism
that screens the background dynamics from the effects of a large coupling can be broken, for relatively large couplings, by the evolution of linear density perturbations.

\section{Linear perturbations}
\label{sec:linear}

We now move to study the evolution of linear density perturbations in the MDM coupled DE scenarios under investigation.
If we define the density contrast of the two CDM species as $\delta _{\pm } = \delta \rho _{\pm }/\rho _{\pm }$, the 
linear perturbation equations for the two fluids are given by (see e.g. \cite{Amendola_2004,Baldi_2011a}):
\begin{eqnarray}
\label{gf_plus}
\ddot{\delta }_{+} &=& -2H\left[ 1 - \beta \frac{\dot{\phi }}{H\sqrt{6}}\right] \dot{\delta }_{+} + 4\pi G \left[ \rho _{-}\delta _{-} \Gamma_{R} + \rho _{+}\delta _{+}\Gamma_{A}\right] \,, \\
\label{gf_minus}
\ddot{\delta }_{-} &=& -2H\left[ 1 + \beta \frac{\dot{\phi }}{H\sqrt{6}}\right] \dot{\delta }_{-} + 4\pi G \left[ \rho _{-}\delta _{-} \Gamma _{A} + \rho _{+}\delta _{+}\Gamma_{R}\right]\,.
\end{eqnarray}
In Eqs.~(\ref{gf_plus}-\ref{gf_minus}) the different signs of the extra friction term (second term in the first squared brackets on the right-hand side) reflect the opposite mass
evolution of the two CDM species with respect to the dynamics of the DE scalar field $\phi $, while the $\Gamma $ factors in the the second squared
brackets are defined as:
\begin{equation}
\label{def_gamma}
\Gamma _{A} \equiv 1 + \frac{4}{3}\beta ^{2}\,, \quad \Gamma _{R}\equiv 1 - \frac{4}{3}\beta ^{2} \,,
\end{equation}
and represent attractive ($\Gamma _{A}$) or repulsive ($\Gamma _{R}$) corrections to gravity due to the
long-range fifth-force mediated by the DE scalar field. As one gets from the definitions (\ref{def_gamma}), a coupling of the order of $\beta \sim 0.15$ 
(which we take as an observational upper limit for standard coupled DE models, see e.g. \cite{Bean_etal_2008,Xia_2009,Baldi_Viel_2010}) determines a correction to standard gravity of the order of
a few percent (i.e. $|\Gamma -1|\approx 10^{-2}$). On the contrary, couplings of order unity and larger might provide a dark scalar force with strength comparable or even larger than gravity, giving rise to very significant effects in the growth of density perturbations. In particular, a coupling of $\beta = \sqrt{3}/2 \approx 0.87$ 
would determine a fifth-force with the same strength as gravity (i.e. $|\Gamma -1| = 1$), thereby resulting in the absence of any force for repulsive corrections ($\Gamma _{R} = 0$), and in a force twice as strong as gravity for attractive corrections ($\Gamma _{A} = 2$). Similarly, a coupling of $\beta =\sqrt{3/2}\approx 1.22$ (i.e. $|\Gamma -1| = 2$) would imply an attractive total force with three times the strength of gravity for attractive corrections ($\Gamma _{A} = 3$), and a repulsive force with gravitational strength for repulsive corrections ($\Gamma _{R} = -1$). 
\ \\

With these definitions, we now consider the evolution of linear density perturbations in matter domination, i.e. when the contribution of perturbations
in the relativistic component of the Universe becomes negligible, and we can therefore include in our discussion only fluctuations 
in the matter sector.

\subsection{Adiabatic and isocurvature modes}

\begin{figure}
\begin{minipage}{70mm}
\includegraphics[width=\linewidth]{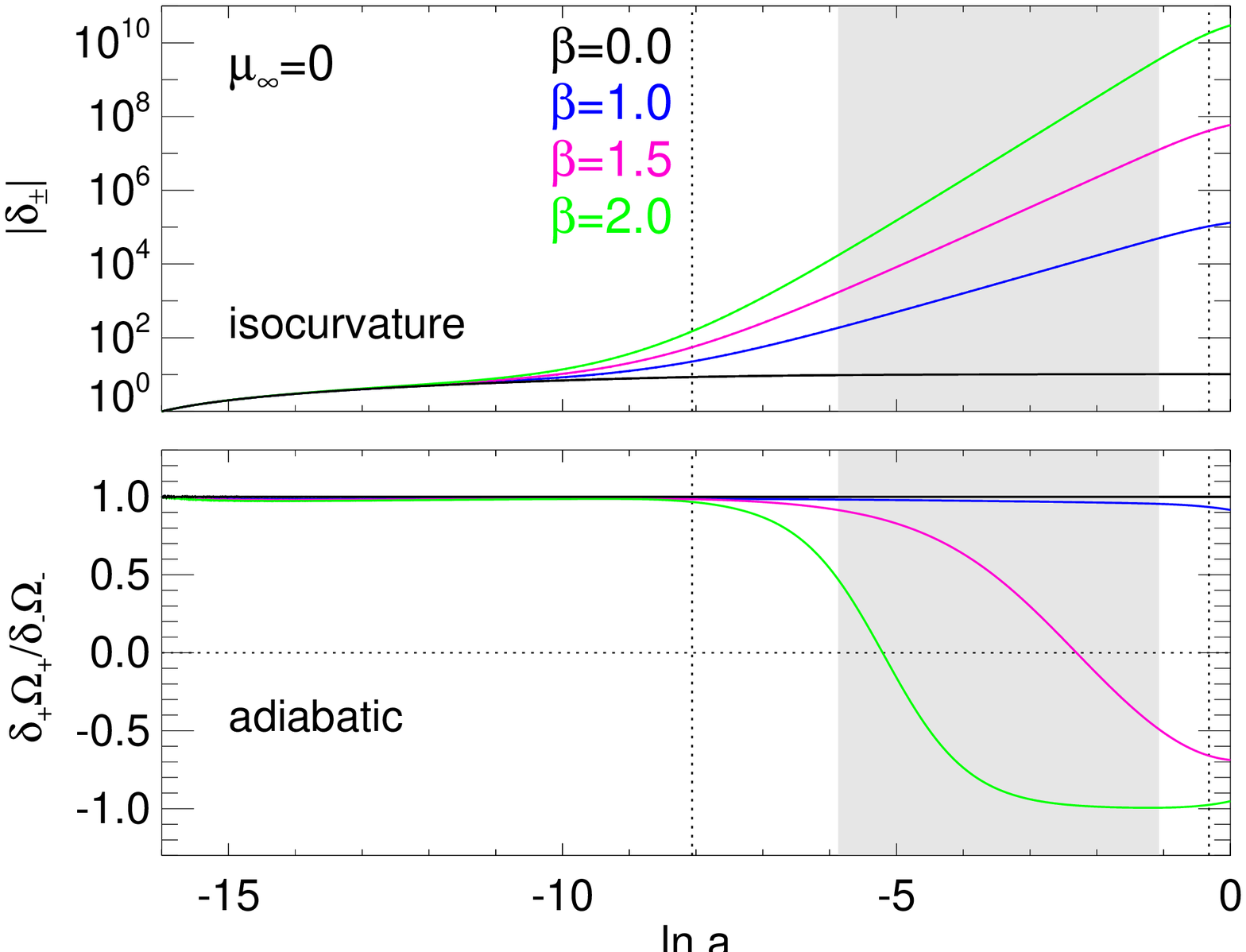}
\caption{{\em Top panel}: the amplitude of density perturbations in the two CDM species for isocurvature initial conditions as a function of the e-folding time. As on can see in the plot, while for the uncoupled case isocurvature modes do not grow for the whole expansion history of the Universe, for MDM coupled DE models the amplitude of isocurvature perturbations grows by several orders of magnitude between the beginning of matter domination and the present time. {\em Bottom panel}: The ratio $\Omega _{+}\delta _{+}/\Omega _{-}\delta _{-}$ as a function of the e-folding time for adiabatic initial conditions. As the plot shows, for the uncoupled case the adiabaticity of perturbations is preserved in time, while for coupled models adiabatic perturbations tend to evolve towards isocurvature.}
\label{fig:iso-adia}
\end{minipage}
\hfil
\begin{minipage}{70mm}
\includegraphics[width=\linewidth]{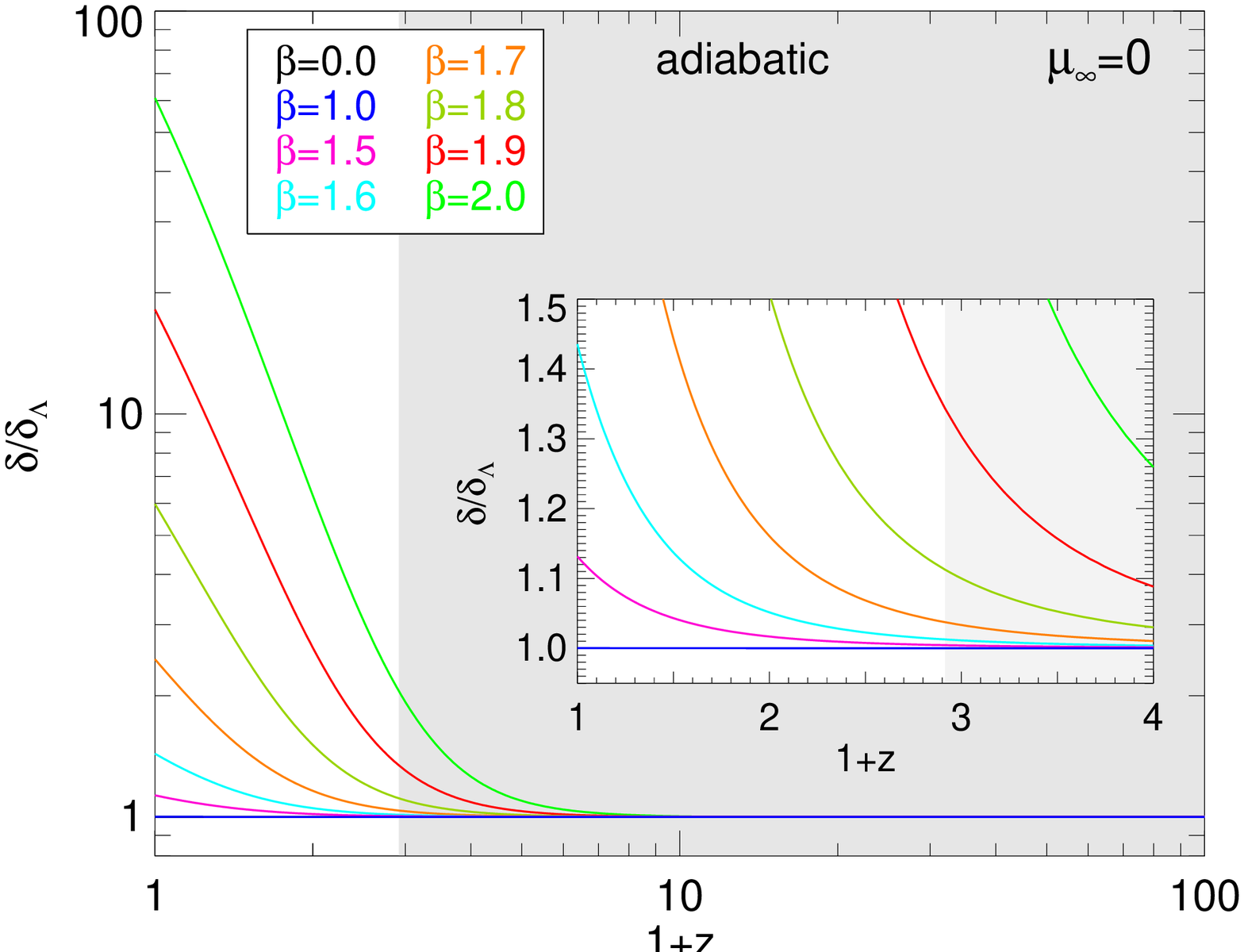}
\caption{The ratio of the growth factor of CDM density perturbations to the $\Lambda $CDM case for a series of MDM coupled DE models with couplings between $\beta = 1$ and $\beta = 2$ and with adiabatic initial conditions. As one can see in the plot, couplings larger than $\beta = 1.6$ (cyan curve) determine an overall enhancement of the amplitude of linear perturbations at $z=0$ exceeding $40 \%$, while a coupling $\beta = 1.5$ (magenta) would induce an enhancement of just $13\%$ and a coupling $\beta = 1$ does not induce any boost in the growth of perturbations, thereby resulting indistinguishable from $\Lambda $CDM. The small box shows a zoom of the same quantities on a linear scale and at low redshifts.}
\label{fig:growthfactor}
\end{minipage}
\end{figure}

For a set of isocurvature perturbations in the matter sector, 
i.e. whenever the condition $\delta_{+}\Omega _{+} = -\delta _{-}\Omega _{-}$ holds, 
the standard gravitational source term of each perturbation equation (\ref{gf_plus}-\ref{gf_minus}) exactly vanishes, and the scalar fifth-force remains the only source term for the evolution of the density perturbations in the two matter species, which will keep growing maintaining their opposite signs, such
that Eqs.~(\ref{gf_plus}-\ref{gf_minus}) become: 
\begin{eqnarray}
\label{gf_plus_iso}
\ddot{\delta }_{+} &=& -2H\left[ 1 - \beta \frac{\dot{\phi }}{H\sqrt{6}}\right] \dot{\delta }_{+} + \frac{32\pi G}{3}\rho _{+}\delta _{+}\,, \\
\label{gf_minus_iso}
\ddot{\delta }_{-} &=& -2H\left[ 1 + \beta \frac{\dot{\phi }}{H\sqrt{6}}\right] \dot{\delta }_{-} + \frac{32\pi G}{3}\rho _{-}\delta _{-} \,.
\end{eqnarray}
In other words, for a superposition of perturbations in the two CDM fluids with opposite density contrast, the overdense species will become progressively more overdense, while the underdense species will become more underdense due to their mutual repulsion. In this respect, we can already qualitatively highlight one peculiar feature of the evolution of linear density perturbations in MDM coupled DE models, i.e. the fact that isocurvature modes do not decay but actually grow in time due to the repulsive nature of the scalar fifth-force between the two different CDM species. This is shown in the upper panel of Fig.~\ref{fig:iso-adia}, where we plot the amplitude of the density perturbations of the two CDM species $|\delta _{\pm}|$ for isocurvature initial conditions normalized to their initial value at $z_{i}\approx 10^{7}$, as a function of the e-folding time. 
As the figure shows, while in the absence of coupling (black curve) the perturbations amplitude remains frozen during the whole
expansion history of the Universe, for progressively larger values of the coupling $\beta $ the amplitude of both the positive 
and negative density fluctuations grows by several orders of magnitude between the beginning of matter domination and the present time. 

\ \\

On the other hand, for adiabatic perturbations, 
i.e. for the case $\delta _{+}\Omega _{+} = \delta _{-}\Omega _{-}$, 
the overall force acting on each of the two density perturbations $\delta _{\pm }$ will just be given by standard gravity since the attractive and repulsive corrections on the right hand side of Eqs.~(\ref{gf_plus}-\ref{gf_minus}) exactly cancel each other, which gives:
\begin{eqnarray}
\label{gf_plus_adiab}
\ddot{\delta }_{+} &=& -2H\left[ 1 - \beta \frac{\dot{\phi }}{H\sqrt{6}}\right] \dot{\delta }_{+} + 8\pi G\rho _{+}\delta _{+}\,, \\
\label{gf_minus_adiab}
\ddot{\delta }_{-} &=& -2H\left[ 1 + \beta \frac{\dot{\phi }}{H\sqrt{6}}\right] \dot{\delta }_{-} + 8\pi G\rho _{-}\delta _{-}\,.
\end{eqnarray}
However, the adiabaticity of matter perturbations (which reduces to the condition $\delta _{+}=\delta _{-}$ for a symmetric state $\mu = 0$)
is bound to be broken by any dynamical evolution of the scalar field $\phi $ as a consequence of the opposite sign of the extra friction terms in Eqs.~(\ref{gf_plus_adiab}-\ref{gf_minus_adiab}), such that even if the initial conditions and the gravitational source terms are the same for the two perturbations, their dynamic evolution 
will be different due to the different friction terms. Therefore, although in matter domination -- as we showed in the previous Section -- the system is attracted towards the uncoupled state $\mu = 0$, and the scalar field is frozen in the minimum of its effective potential $V_{\rm eff}$, any oscillation around this minimum will induce a departure from adiabaticity of any initially adiabatic set of density perturbations, and will consequently restore the 
fifth-force corrections $\Gamma $ in the evolution equations ~(\ref{gf_plus}-\ref{gf_minus}). Then, for coupling values $\beta \geq \sqrt{3}/2$, once the fifth-force is no longer suppressed by the adiabaticity of the perturbations, the overdensities in the two different matter species will start repelling each other and only the fluctuation with the (even slightly) larger amplitude will keep growing, while the other will start slowing down its growth and then decay, thereby moving the system towards isocurvature. We can therefore conclude that the adiabaticity of density perturbations is an unstable condition for MDM coupled DE models, and that any adiabatic set of perturbations will evolve towards isocurvature due to the instability generated by the extra friction terms in Eqs.~(\ref{gf_plus}-\ref{gf_minus}).
This evolution, which represents another distinctive feature of MDM coupled DE scenarios, is shown in the bottom panel of Fig.~\ref{fig:iso-adia}, where we plot the quantity $\delta _{+}\Omega _{+}/\delta _{-}\Omega _{-}$
as a function of the e-folding time for adiabatic initial conditions. This ratio quantifies the
level of adiabaticity of the density perturbation with $\delta _{+}\Omega _{+}/\delta _{-}\Omega _{-}=+1$ corresponding to a pure
adiabatic mode, and $\delta _{+}\Omega _{+}/\delta _{-}\Omega _{-}=-1$ to an isocurvature mode. As one can clearly see in the plot,
in the absence of coupling (black curve) adiabatic modes remain adiabatic during the whole expansion history of the Universe.
On the other hand, for MDM coupled DE models with $\beta > 0$ we can clearly see how an initially adiabatic set
of perturbations evolves in time towards isocurvature. The effect is proportional to the coupling strength $\beta $ and
while still relatively modest for couplings of order $\beta \approx 1$ it becomes more significant for $\beta \gtrsim 1.5$ with the ratio 
$\delta _{+}\Omega _{+}/\delta _{-}\Omega _{-}$ reaching a value of $-0.7$ at $z=0$ for $\beta = 1.5$ and even approaching $-1$ already at high redshifts for $\beta = 2$.

\subsection{Linear growth for symmetric models with adiabatic initial conditions}

We now restrict our attention to the specific case of symmetric models (i.e. models with $\mu _{\infty } = 0$)
with adiabatic initial conditions, which represent the most realistic situation for our proposed
scenario. For this setup we investigate the dynamics of the total linear density perturbations defined as:
\begin{equation}
\delta _{\rm CDM} \equiv \frac{\Omega _{+}\delta _{+}}{\Omega _{\rm CDM}} + \frac{\Omega _{-}\delta _{-}}{\Omega _{\rm CDM}} \,
\end{equation}
by numerically solving Eqs.~(\ref{gf_plus}-\ref{gf_minus}) for different values of the coupling $\beta $ along the corresponding background evolution. The results of such integration, which represent the total linear growth factor of CDM density perturbations in our MDM coupled DE models, are shown in Fig.~\ref{fig:growthfactor}, where we plot on a log-log scale the ratio of the perturbations amplitude $\delta _{\rm CDM}$ over
the $\Lambda $CDM case $\delta _{\Lambda }$ (black curve, corresponding to the uncoupled model $\beta = 0$) for a large number of
coupling values between $\beta = 1$ and $\beta = 2$ as a function of redshift. The small plot in Fig.~\ref{fig:growthfactor} shows a zoom
of the same quantities at $z<3$ on a linear scale and for $\delta _{\rm CDM}/\delta _{\Lambda } < 1.5$ in order to allow an easier inspection of the results for small coupling values. In both plots the grey-shaded area indicates FMD. As one can see in the figure, the total growth
of CDM density perturbations in MDM coupled DE models significantly deviates from the $\Lambda $CDM case at low redshifts ($z\lesssim 10$) reaching at $z=0$ an amplitude enhancement of a factor $\sim 60$ for $\beta = 2$, while at higher redshifts the evolution remains
indistinguishable from $\Lambda $CDM. Such fast growth of linear density perturbations at low redshift would result in a huge mismatch
between the value of $\sigma _{8}$ measured from local probes and the value inferred from the amplitude of scalar perturbations
at last scattering under the assumption of a standard $\Lambda $CDM cosmology. More specifically, for $\beta = 2$ one would measure 
the unrealistic value of $\sigma _{8}\approx 50$ today while having the same normalization of the CMB quadrupole as in a $\Lambda $CDM scenario. 
Such value of the amplitude of linear density perturbations is obviously starkly incompatible with even the most basic observations of the cosmic Large Scale Structure, and clearly rules out the model.
The evolution of density perturbations therefore allows in principle, as suggested also by BVH08 for their more general scenario, to rule out MDM coupled DE models that would be otherwise considered perfectly viable from their background expansion history: if at the background level, as we showed in 
Section~\ref{sec:background}, couplings as large as $\beta = 10$ would appear perfectly acceptable for a MDM coupled DE scenario,
this is no longer true for linear perturbations where a coupling of $\beta \sim 2$ can be easily disproved.
In this respect, then, our results broadly confirm the previous findings of BVH08. 

However, our range of parameters for the specific realization
of MDM coupled DE scenarios discussed in this work is significantly larger than in BVH08, and allows us to investigate in more detail 
to which extent linear density perturbations do really provide a way to break the degeneracy between MDM coupled DE models
and uncoupled cosmologies that was shown to hold at the background level for arbitrarily large values of the coupling $\beta $.
When looking at Fig.~\ref{fig:growthfactor}, in fact, one can notice that the effect of enhanced growth strongly depends on
the coupling itself: if coupling values larger than $\beta = 1.6$ (cyan curve) can be immediately ruled out as they would imply $\sigma _{8}(z=0)\gtrsim 1$, smaller couplings in the range $\beta \lesssim 1.5$ appear still viable also at the linear level, as the predicted
value of $\sigma _{8}$ does not significantly exceed $\sigma _{8}(z=0)\lesssim 0.9$. In particular, it is very interesting to notice that a coupling of $\beta =1$ does not show any enhancement at all and features -- besides the background evolution -- also a growth of linear density perturbations completely indistinguishable from $\Lambda $CDM. Couplings of order unity therefore cannot be ruled out even at the linear level in MDM coupled DE models, and this is of course equally true for even smaller couplings $\beta \lesssim 1$. 

In this respect, our study therefore limits the validity of the claim of BVH08 that linear perturbations allow to distinguish MDM coupled DE from an uncoupled cosmology only to relatively large coupling values.  On the contrary, our work shows for the first time that a significant portion of the parameter space of MDM coupled DE models -- that is ruled out for standard coupled DE with one single CDM species -- turns out to be viable both at the background and at the linear perturbations level. Therefore, we have proven here that linear probes are not in general sufficient to rule out long-range scalar interactions of gravitational strength in the dark sector. It is then natural to speculate whether extending the analysis to the nonlinear regime of structure formation could further reduce the allowed parameter space for these scenarios. Such analysis is left for future work.

\section{Conclusions}
\label{sec:concl}

In the present paper we have studied in detail the background and the linear perturbations evolution
of cosmological models featuring two different species of CDM particles interacting with opposite coupling constants
with a classical scalar field responsible for the observed accelerated expansion of the Universe. Such models represents a specific realization of the more general framework proposed by {\em Brookfield, van de Bruck \& Hall, 2008} 
that allows to reduce the parameter space of such more general scenario to the same dimension
of a standard interacting dark energy cosmology. 
\ \\

For this class of models we have studied in detail the background evolution starting from the same initial conditions of a minimally coupled scalar field cosmology with an expansion history indistinguishable from the concordance $\Lambda $CDM scenario. Our analysis has shown that the presence of two different CDM species interacting with
opposite couplings with DE provides a very effective self-regulating mechanism that screens the background evolution of the Universe from arbitrarily large values of the coupling strength. More specifically we have shown, 
confirming previous results, that such MDM coupled DE models feature an expansion history practically indistinguishable from $\Lambda $CDM even for coupling values as large as $\beta =10$. Furthermore, extending
previous investigations, we have studied how this self-regulating mechanism depends on the initial conditions
of the system, in particular on the relative abundance of the two CDM species at high redshifts. In this respect, we found that
a large asymmetry between the two CDM particle types in the early Universe could significantly reduce the
efficiency of the screening and determine expansion histories clearly incompatible with observations. However, 
we also found that a primordial asymmetry of about 10\% does not significantly weaken the effectiveness of the screening and that therefore no real fine-tuning of the primordial relative abundance of the two CDM species
is required in order to provide viable background solutions even for large coupling values.
\ \\

We have then studied the evolution of linear density perturbations in the context of such MDM coupled DE
scenarios. In particular, we focused on the evolution of isocurvature and adiabatic perturbations modes
in the matter sector, showing how, differently from what happens in $\Lambda $CDM as well as in standard coupled DE models with one single CDM species, isocurvature perturbations significantly grow in time during matter domination for sufficiently large couplings due to the repulsive long-range fifth-force between density fluctuations in the two different CDM fluids. Furthermore,
we have also shown how starting from an initial set of adiabatic perturbations these evolve in time towards isocurvature in MDM coupled DE scenarios. This peculiar behavior is not realized neither by minimally coupled cosmologies nor
by standard coupled DE models with one single CDM species, and therefore represents a clear distinctive feature
of these scenarios. Finally, we have investigated the evolution of the total CDM density perturbations for models
with adiabatic initial conditions, finding that for sufficiently large values of the coupling the growth rate is enhanced
at low redshifts with respect to $\Lambda $CDM.

This shows that the evolution of linear density perturbations can in principle break the screening mechanism of MDM coupled DE that protects the background evolution of the Universe even from extremely large values of the dimensionless coupling $\beta $. Therefore, tests of the linear growth allow in principle to distinguish between a coupled and an uncoupled cosmology, in two ways:\\
\begin{itemize}
\item[$i)$] the emergence of isocurvature modes even from an initial set of adiabatic density perturbations in the matter sector represents 
a clear distinctive feature of MDM coupled DE models, and might provide a direct way to test and constrain the scenario; similarly,
since isocurvature perturbations are expected to grow in time in these models, present constraints on the amount of primordial isocurvature modes from CMB observations could be directly turned into constraints on the allowed parameters range for this scenario;
\item[$ii)$] even more importantly, the overall linear growth of the combined CDM density perturbations is strongly enhanced at low redshifts as compared to the
standard $\Lambda $CDM case, such that -- starting from the same normalization of scalar perturbations at last scattering -- MDM coupled DE models predict a value of $\sigma _{8}$ at the present time that can significantly exceed the upper limit allowed by low-redshift observations, thereby providing a direct way to rule out a large portion of the parameter space of the model.
\end{itemize}

In this respect, our results qualitatively confirm previous outcomes on more general realizations of interacting DE models with multiple CDM families.
However, by significantly extending the the parameter range explored in previous works, our analysis has allowed to show how both these effects emerge in a clear way only for relatively large couplings, $\beta \gtrsim 1.6$, while for smaller couplings
the evolution of linear perturbations seems still compatible with present observational bounds. In particular, for couplings of order unity and smaller, both the background evolution and the growth of linear perturbations are completely indistinguishable from the standard 
$\Lambda $CDM case. Therefore, there is a significant range of coupling values, namely  $0.15 \leq \beta \leq 1$, that are ruled
out for standard coupled DE models with a single CDM species, but that appear still perfectly viable (at least up to linear order) if we assume a MDM coupled DE scenario.
The claim made in previous works that linear probes allow to distinguish between an uncoupled cosmology and a MDM coupled DE scenario
is therefore shown by our analysis to be true in practice only for relatively large coupling values.
It is in fact important to recall that a coupling as large as $\beta \sim 1$ implies a scalar fifth-force stronger than standard gravity, and 
therefore determines an overall repulsive interaction between CDM particles of the two different species. Such repulsive long-range interaction
is expected to have significant effects on the dynamics of collapsed structures at small scales that are not well described by linear perturbation theory. In order to investigate such effects and devise new possible ways to constrain MDM coupled DE models
even for couplings of order unity and smaller
 it would then be necessary to extend the analysis to the nonlinear regime of structure formation by means of specific N-body simulations. This goes beyond the scope and the time constraints of the present work, and will be investigated in a separate publication.

\begin{acknowledgement}
I am deeply thankful to Luca Amendola for useful discussions.
This work has been supported by the DFG Cluster of Excellence ``Origin and Structure of the Universe'' and by
the TRR33 Transregio Collaborative Research Network on the ``Dark
Universe''.

\end{acknowledgement}

% Use this code if you wish to generate your bibliography with BibTeX;
% please replace first the string "demo" below with the name(s) of
% the BibTeX data base(s) you want to use.
% The resulting bibliography-output (the contents of the .bbl file)
% must be pasted into this file before submission.
% 
\bibliographystyle{adp}
\bibliography{baldi_bibliography}

\end{document}